# Transmitting signals over interstellar distances: Three approaches compared in the context of the Drake equation

Luc ARNOLD, Aix Marseille Université, CNRS, OHP (Observatoire de Haute Provence), Institut Pythéeas, UMS 3470, 04870 Saint-Michel-l'Observatoire, France

Correspondance: Luc ARNOLD, CNRS - Observatoire de Haute Provence, 04870 Saint-Michel-l'Observatoire, France.  Email: Luc.Arnold@oamp.fr

Abstract: I compare three methods for transmitting signals over interstellar distances: radio transmitters, lasers and artificial transits. The quantitative comparison is based on physical quantities depending on energy cost and transmitting time $L$, the last parameter in the Drake equation. With our assumptions, radio transmitters are the most energy-effective, while macro-engineered planetary-sized objects producing artificial transits seem effective on the long term to transmit an attention-getting signal for a time that might be much longer than the lifetime of the civilization that produced the artifact.

# 1. Introduction

One of the most effective methods to discover exoplanets is the so-called *transit method:* When a planet transits in front of its star, it blocks a small fraction of the stellar light, producing a small decrease of the observed stellar flux. The depth of the transit light curve is proportional to planet cross-section. Assuming the *shape* of the transiting object is *spherical*, which is a reasonable model for a planet, then the object cross-section is just proportional to the *planet radius* squared. What happens if the object is *not spherical?* It has been shown that the transit light curve contains a signature of the transiting object shape: For a given cross-section, if the planet is ringed (Saturn-like) or slightly flattened by rotation, the shape of the transit light curves will indeed slightly differ from the transit of a spherical body (for example Barnes & Fortney, 2003; Barnes & Fortney, 2004). In fact, many parameters of the planet/star system can be (at least in principle) derived from the transit light curve (Winn, 2008).

In 2005, I proposed that, in our current successful and growing research of transiting exoplanets with the KEPLER (Bathala et al., 2013) or Corot (Rouan, 2012) missions, we might detect the transits of artificial planetary–sized objects in orbit around stars to produce attention-getting signals from other civilizations (Arnold, 2005a).

Looking for artificial transits offers another Dysonian approach to SETI (Search for Extraterrestrial Intelligence). Dysonian SETI aims to look for signatures of macro-engineering activities in space (Cirkovic, 2006; Bradbury et al., 2011). The concept originates with Dyson (1960) who suggested to look for infrared radiations excess from stars that could be the signature of degraded energy leaking from a spherical structure built around a star to trap stellar energy and feed an advanced civilization (Bradbury, 2001). Sagan and Walker (1966) concluded that such objects should indeed be detectable, although probably difficult to distinguish from natural low-temperature objects. Slysh (1985), after an analysis of IRAS (Infrared Astronomical Satellite) data, had the same conclusion and therefore proposed to extend the search to micro- and millimeter wavelengths. At least Slysh (1985), Jugaku & Nishimura (2004) and Carrigan (2009) carried searches for Dyson spheres. These surveys, or other surveys by Annis (1999), also belong to Dysonian SETI activities. Harris (1986) proposed another Dysonian SETI, considering the detectability of civilizations 'burning' antimatter, which should radiate in the $\gamma$-ray spectrum. He made a survey of $\gamma$-ray observational data to look for such a signature and was able to put an upper limit on that activity (Harris, 2002). The search of extraterrestrial artifacts (SETA) or extraterrestrial technologies, not necessarily Dysonian, have been proposed or reviewed in several papers (Bracewell, 1960; Papagiannis, 1978; Freitas, 1980; Freitas, 1983; Freitas and Valdes, 1985; Ellery et al., 2003). Clearly, all these alternative - or *unorthodox* as already qualified by Harris in 1986 - approaches allow the exploration of the widest spectrum of possibilities and are keys for the desirable success of SETI. As Freeman Dyson says, "look for what's detectable, not for what's probable"[1].

Moreover, although artificial transit search falls into that *unorthodox* approach, it remains in full agreement with Tarter's assumption regarding quasi-astrophysical signals (Tarter, 2001) for SETI: "An advanced technology trying to attract the attention of an emerging technology, such as we are, might do so by producing signals that will be detected within the course of normal astronomical explorations of the cosmos." Radio or transit signals both fit this statement.

---

[1] F. Dyson in February 2003, in a TED conference in Monterey, California (http://tedxproject.wordpress.com/2010/05/23/freeman-dyson-lets-look-for-life-in-the-outer-solar-system/).

In this paper, I will compare artificial transits for transmitting signals with radio broadcasting as proposed by Cocconi and Morrison (1959) and Drake (1960). I will also consider laser pulse transmission as proposed by Schwartz and Townes (1961) and Ross (1965). On the *listener* point of view, the classic SETI approach (Tarter 2001) looks for radio signals, while optical SETI aims to detect laser pulses sent by extra-terrestrial beings. Several optical SETI projects are or were operated (Kingsley, 2001; Howard et al., 2004; Mead and Horowitz, 2010).

I will emphasize on effective methods for transmitting attention-getting signals rather than for communicating complex messages. Thus the inscribed-matter approach (Rose and Wright 2004), although effective (if haste is unimportant, according to the authors) to deliver very large amount of informations ($10^{22}$ bits/kg) at a given place, is not effective for an attention-getting signal, where only a few bits are sufficient.

The Drake equation (for example Ulmschneider, 2003) allows evaluating the number $N$ of communicating civilizations in our galaxy. This number is calculated as a product of different probabilities of which several remain mostly unknown. The number $N$ varies by several orders of magnitude depending on the authors, underlining the weak predictive behaviour of the equation with the current weakly constrained *theoretical probabilities* in the equation. The observation and statistics of exoplanets might help to derive in the near future an *observational frequency (or probability)* for planetary systems that are suitable for life, but other, such as the probability that live evolved toward a communicative civilization will remain unknown, even after a first contact with an alien civilization that will just tell us that that probability is not zero (although this will undoubtedly be a fantastic information). The Drake equation is nevertheless recognized as an effective tool to identify the relevant parameters to take into account when *we think about N*.

If we consider $N$ as the number of communicating civilizations in our galaxy that we might detect at a given time, we understand that $N$ will increase if civilizations transmit a signal for a long time. A civilization wanting to communicate (or just to be detected/seen) will indeed maximize the time $L$ during which it transmits a signal. The transmitting time $L$, the last factor in the Drake equation, is also sometimes considered as the lifetime of the civilization supposed to transmit information continuously as soon as it has the technology to do so.

In the following sections, I will compare the relevance of SETI at radio wavelength, optical SETI and artificial transits with respect to $L$. I will do so by considering *the listener position* (us doing SETI), but I will especially emphasize the discussion on arguments relevant for the *broadcasting side*. I understand that it is not easy to compare existing technologies like radio and lasers with speculative technology like artificial transit. I will try to take particular attention to this aspect in the following discussion.

In Section 2, I define some physical quantities related to the transmission of signals with the mentioned methods and evaluate their values. These values are discussed in Section 3.

# 2. Relevant physical quantities to compare communication methods

We consider now physical quantities related to energy cost and transmitting time, two relevant variables for the transmission of signals. The quantities considered below do probably not form an exhaustive list.

## 2.1. Time to transmit a signal toward 50000 stars

Let us define the period $P$ to transmit a signal once toward a given number of stars. Let us consider a transiting object at 1 AU from a solar-type star. The solid angle over which the transit (here the signal) is transmitted over one year is 0.5% of the full sky (Arnold, 2005a). How many stars are reached by the transit signal? The stellar density in the Sun vicinity is known to be about 0.13 $pc^{-3}$ (Bahcall, 1986), but this density is rather valid within 5 pc around the Sun, and falls to 0.046 $pc^{-3}$ within a sphere of 25 pc radius (Jahreiss et al.,1999; Robin et al., 2003). In the same sphere around the Sun, Raghavan et al. (2010) counted 454 dwarf and subdwarf solar-type stars (~F6-K3 spectral types), a number therefore representing 15% of the total number of 2996 stars within 25 pc (Jahreiss et al., 1999). Extrapolating the validity of these stellar densities to a sphere of 300 pc leads to a total number of stars between 5 and $15.10^6$. It means that typically 25 000 to 75 000 stars, of which 4000 to 11 000 are solar-type stars, are reached by the transit attention-getting signal during $P=1$ $yr$. In the following, let us consider a mean number of targets of 50 000 stars including 7500 solar-type stars. A sphere of 300 pc is chosen here because 1) it contains the targets the radio Allen Telescope Array permits to explore (Turnbull and Tarter 2003); 2) it is the volume considered by Kingsley (2001) for optical SETI; 3) at a distance of 300 pc, solar-type stars have apparent magnitudes of V=19.7 for a 0.1 solar luminosity K3 subdwarf, V=17.2 for a Sun twin, or V=14.7 for a 10 solar luminosity F6 star. These magnitudes allow the photometry at 1% or even better accuracy required for the search for transits.

Let us also consider that a transiting object, if likely built in the habitable zone near the home planet at $a=1$ (where $a$ is the semi-major axis), can be injected, once completed, into an orbit closer to the star in order to increase both transit frequency, varying as $a^{-3/2}$, and solid angle $\Omega$ over which the transit is transmitted. The solid angle $\Omega$, or the number of stars reached, is proportional to $1/a$ (Arnold, 2005a). For an object on a Mercury-like orbit ($a=0.4$ AU), the solid angle is increased by a factor 2.5, and the orbital period $a^{3/2}$ divided by a factor of 4. A total of more than $10^5$ stars should be reached in principle in a complete orbit of 0.25 $yr$, and 50000 stars would be reached in only $P=0.1$ $yr$, with transits still lasting about 6 hours each. For an object at 1 AU, a transit typically lasts 10 hours - up to 13 hours if the impact parameter is near zero - during which the signal can be modulated for example by the rotation of a non-spherical object to be recognized as artificial.

If one considers the transmission of a radio signal toward 7500 solar-type stars with Arecibo-size (or smaller) antennas, it is likely that all stars could in principle be targeted once (one visit for one star) within about one year too, at a rate of about one target per hour. In order to spent 10 hours per star as with a transit, 10 times more antennas would be needed, or of the order of 100 times more to reach 50000 stars for 10 hours each, otherwise of the order of 100 $yr$ would be needed.

For the transmission of a laser pulse toward the same number of stars, the same order of time would be needed in principle. Nevertheless, as already pointed out (Arnold, 2005a), it is necessary to accurately know the proper motion of the target star otherwise the (visible or near infrared) laser beam will miss its target[2]. Without this knowledge, a larger patch of sky has to be targeted, increasing the number of pulses to be sent by a factor up to $10^6$. This factor is the ratio of the sky surface to be targeted (a few square-arcsec) over the telescope resolution squared (0.01 arcsec for a 10-m telescope). The pulse frequency needs to reach at least tens of Hz if one hour of broadcasting per star is planned, with the perspective that only one pulse will reach the target. This shot frequency is yet far from current human technology: The Laser Mégajoule LMJ (André, 1999) should allow only 30 shots/year at full power ($6 \times 10^{14}$ W). It is thus a waste of time and energy to transmit without an accurate knowledge of the target proper motion. The accuracy required is $10^{-5}$ arcsec/year if the beacon is transmitted with a 10-m telescope on a target at 300 *pc* that would move by 10 arcsec in 1000 *yr*. But let us assume in the following that this astrometric accuracy is reached. To transmit towards 7500 stars at about one hour per star, again of the order of one year is needed. In order to spent 10 hours per star as with a transit, and as with radio broadcasting, 10 times more lasers would be needed, or of the order of 100 times more to reach 50000 stars, otherwise of the order of 100 *yr* would be needed.

The conclusion of this first sub-section is that radio or laser transmission would reach the same number of stars in *P*=1 to 100 *yr* that the transit method would in 1 *yr*. There is a constraint of targets prioritisation that does not exist with transits. We report all *P* values in the first line of Table 1.

## *2.2. Energy to transmit a signal toward 50000 stars*

For transits, the light of the transited star carries the signal. Therefore the energy cost is basically the energy invested in the building of the transiting object. The energy needed to produce a thin mask of iron of the diameter of the Earth has been evaluated to be of the order of $10^{15}$ Wh, about 3 days of humanity energy consumption in the 2000's (Arnold, 2005b). Let us consider that the investment is about $E_p=10^{16}$ Wh for some masks to produce recognizable artificial multiple transits (Arnold, 2005a). We consider that once the transiting object is operational, it will not require (significant) additional energy to transmit.

The power of the Arecibo radar is $10^6$ W. The Yevpatoria radar is in the $10^5$ W range. The energy needed to transmit during 10 hour/star (as for transits) toward 50000 stars is thus of the order of $E_p=10^{11}$ Wh.

The laser invoked by Kingsley (2001) produces nanosecond pulses reaching $10^{18}$ W, with a duty cycle of $10^{-9}$. Kingsley's laser mean-power is thus $10^9$ W. The energy used to transmit during 10 hour/star (as for transits) toward 50000 stars is about $E_p=10^{14}$ Wh. Pulse 100 times less powerful, yet be detectable at 300pc, would give $E_p=10^{12}$ Wh.

It can be argued that for radio transmission, the signal could be fractioned into pulses like for lasers. Ross (2000) suggests that laser pulses should be separated by more than 10s. We can thus propose that radio signal could be transmitted for 100ms every 10s for 10h for each targeted star. This would lead to $E_p=10^9$ Wh for a radio transmitter. And for laser, with one pulse every 10s instead of 1s, $E_p=10^{11}$ Wh.

---

[2] This is of course also true in principle for radio beaming, but not relevant in practice, because the patch of the beam at radio wavelengths is of the order of 2 *arcmin* (transmission at wavelength $\lambda$=21cm with the Arecibo dish for example), which is much larger than the targets proper motion range.

Note that we may speculate that the power of aliens' transmitters could also be much more powerful than the Arecibo or Yevpatoria radars, but since the power signals of these existing radars can be detected at 300pc by similar antennas, let us consider that more power affords no advantage within the assumptions discussed here. I report these energy values in Table 1.

### *2.3. Time of transmission*

As said in the Introduction, to maximize the chance of establishing contact, a good way is to maximize the time of transmission *L*. Transmission technologies that need a continuous feed of large energy quantities will depend on the reliability of maintenance availability in the long term, and thus will stop if the civilization disappears. *L* may thus not exceed the lifetime of a civilization. But we cannot preclude in principle that a reliable stand-alone power plant, able to transform one of the primary energy sources on the planet (sun, wind, water and tides, geothermal sources) into electricity may exist in the future. This would allow transmitting eventually after the civilization has disappeared, to leave a trace of its *passage*. But let's consider that for radio and laser transmission, *L* is limited to the civilization lifetime estimated in the range 100 - 1000 *yr*.

The transmission time with transits is related to the lifetime on the transiting object itself rather than the lifetime of the civilization. If we consider small-mass objects like asteroids in the vicinity of the Earth orbit, the stability of their orbit is of the order of $10^5$ to $10^6$ *yr* (Michel, 1997; Connors et al., 2011; Dvorak et al., 2012;). A particle may even be trapped in a horseshoe orbit for $10^9$ *yr* in the Sun-Earth system if we consider the lifetime of such orbits defined in Dermott and Murray (1981), although other interactions may reduce this extreme duration. Scholl et al. (2005) also established that Venus Trojan orbits can be stable over $10^8$ *yr*. The values for *L* are reported in Table 1.

### *2.4. Energy invested per year of transmission*

The energy per year of transmission, or the annual power invested for the project, is given by $E=E_p/P$, which can also be written as $E=E_p \times n / L$ where $n = L/P$ is the number of visits of one star during the transmission program. For transit, we have $E=E_p/L$. The values for E are reported in Table 1.

Table 1: Values of physical quantities related to the signal transmitters. The number in first column refers to the sub-section in the text where the quantity in the other column is discussed.

|  | Radio | Laser | Transit |
|---|---|---|---|
| 2.1 Time to transmit once toward 50000 stars, $P$ [yr] | 1 … 100 | 1 … 100 | 1 ($a$=1AU)<br><br>0.1 ($a$=0.4 AU) |
| 2.2 Energy to transmit once toward 50000 stars, $E_p$ [Wh] | $10^9$ … $10^{11}$ | $10^{11}$ … $10^{14}$ | $10^{16}$ |
| 2.3 Transmission time $L$ [yr] | 100 … 1000 | 100 … 1000 | $10^3$ … $10^8$ |
| 2.4 Energy invested per year of transmission, or annual power, E = $E_p/P$ [Wh/yr] or E= $E_p/L$ for transits | $10^7$ … $10^{11}$ | $10^9$ … $10^{14}$ | $10^8$ … $10^{13}$ |
| 3. Energy invested per year of transmission and per star (50000 stars), or annual power per star $E_s$ [Wh/yr/star] | $10^2$ … $10^6$ | $10^4$ … $10^9$ | $10^3$ … $10^8$ |

## 3. Discussion

Table 1 and the assumptions presented above show that a radio transmitter is the most efficient for a short program, while lasers may require $10^2$ times more energy. Transits become interesting only on the long term. They might thus be used by advanced civilizations (probably much older than us as suggested by the transmission technology used), energy-rich, to produce attention-getting signals, and/or motivated to leave a trace of its passage in the galaxy. I therefore suggest that, if artificial transits are detected, they might be interpreted as a message of an old and stabilized civilization, or as a volunteer trace of a defunct civilization. Without more informations, it will nevertheless not be possible to decide if it is a signature of a xenoarcheologic object or not (McGee, 2010). At least, it would indicate that the lifetime of a technological civilization could be larger than several centuries, much more than our 100-*yr* young technological and not stabilized civilization. I note that our civilization already managed the amount of energy needed for the construction of transiting objects (Arnold, 2005b), although we do not yet have the technology to do it.

Could the energy $E_s$ invested per year of transmission and per star (bottom line in Table 1) be decreased for transits, which would make transits more efficient on a shorter term? In section 2.1, I point out that injecting an object from $a=1$ to 0.4 AU would decrease the transit duration by 37% (which varies as $\sqrt{a}$) but increase the frequency of the transits by a factor of 4 (transit frequency varying as $a^{-3/2}$). The transmission time thus varies as $a^{-3/2}\sqrt{a} = 1/a$, meaning that at 0.4 AU, the transmission (transiting) time toward a given star is increased by 1/0.4=2.5 with respect to an orbit at 1 AU. Thus, not only $E_s$ would decrease as $1/\Omega$ or $a$, but also the transit (transmission) time $1/a$ for a given target star. But to gain an order of magnitude on $E_s$, the semi-major axis $a$ should be 0.1 AU, a possibly too close distance to the star.

If the orbit of the transiting object is in a plane close to the galactic plane, more stars per year (per orbit) can be reached. An inclination with respect to the galactic plane between 10 … 20° (to avoid too much star crowding in the galactic plane) would allow to reach about $10^6$ stars brighter than R=17.5 (Bahcall, 1986) per year. The value for $E_s$ would decrease by an order of magnitude with respect to the number of stars considered in the first assumptions (50000). If we consider that radio or laser transmitters could also reach these $10^6$ stars within their lifetime, $E_s$ would decrease too, but this would mean that the transmitting time per star would decrease in the same ratio too. If the number of transmitters is multiplied, then the transmitting time per star would be maintained, but $E_s$ would remain unchanged.

The sky coverage with transits is increased too if the object is placed on an orbit around a planet with a significant precession of the nodal line (orbits in relatively high inclination - perhaps 30-45 degrees or more). For the Moon, the precession of this line occurs in 18.6 *yr*. Such orbits increase the sky-coverage for transmitting attention-getting signals on the long term. For an object in orbit around a planet, like the Moon around the Earth, the stars targeted year after year would change over a period of 18.6 *yr*.

The sky coverage with transits is increased again if the object oscillates in tadpole or horseshoe orbits with respect to a planet and if orbit inclination is > 0°, as for Earth Trojan asteroid 2010 TK7, or also Earth co-orbital asteroid 3753 Cruithne. The period of this oscillation is 395 *yr* for 2010 TK7 (Connors et al., 2011) and typically several centuries for other objects.

At last, we may speculate that, at some point, it would even be possible to use the Lidov-Kozai mechanism to oscillate between orbits with different inclination and eccentricity to maximize the visibility of transits on very long periods of time. The discussion above about sky coverage with transits is speculative, and other dynamical effects should be taken into account (Yarkovsky effect (Bottke et al., 2002), etc.) to assess the relevance of the above propositions.

On the *listener* point of view, Table 1 confirms the relevance of SETI at radio wavelengths. It has been pointed out (Kilston et al., 2008) that, because extraterrestrial civilizations in the ecliptic plane would see the Earth transiting the Sun, it is relevant for SETI to search along the ecliptic line. This is also relevant for optical SETI, transit searches and all kinds of Dysonian SETI. The Kepler mission monitors $10^5$ stars over $10^2$ square-degrees (0.24% of the full sky, and far from the ecliptic plane) for transits and already announced more than $10^3$ transit candidates, among which tens have become confirmed planets. Kepler, Corot, or other ground-based programs are the first to explore the sky for transiting planets, and, as for radio wavelengths or optical SETI, we can consider that we are at the very beginning of the search for transiting planets – and artificial transits.

We may also argue that a civilization wanting to communicate with other beings also may want to leave a trace or an artifact in the galaxy that would survive much longer that the civilization itself. These two civilization behaviours seem not incompatible, but rather naturally linked and complementary, at least from an anthropocentric psychological point of view. But artificial planetary-sized objects may also be built for other technological purposes than communication, like energy gathering for example. Such macro-engineering achievements could be the result of natural technological evolution (Dick, 2003; Dick, 2008) making the will or desire of communication only an optional argument.

The number $N$ of communicating civilizations, either alive or extinct, in our galaxy that we may observe today is equal to the sum (integral) of all technological civilizations that emerged in the Galaxy since its birth, assuming all civilizations left a trace somewhere that survived in the Galaxy for more than $10^{10}$ $yr$. Neither radio or laser transmission, nor transits apparently may survive this long. But since transiting objects can survive much longer than other active techniques, this reasoning shows that $N$ increases if we consider artifacts able to survive (transmit) over a very long time.

## 4. Conclusion

In the analysis above, I wanted to emphasize that the last parameter $L$ in the Drake equation, often presented as the civilization lifetime, or in a SETI perspective, being the transmission time of that civilization, is a key parameter in order to compare different transmitters for interstellar communication and thus SETI approaches. Based on quantitative assumptions, I have shown that SETI at radio wavelengths is relevant in term of energy cost. Transmitting through artificial transits becomes energy-efficient if the transmitting time can be very long, possibly much longer than the lifetime on the civilization itself. I therefore suggest that if such artificial transits are detected, they might be the signature of *i)* a defunct civilization, which nevertheless reached a higher technological level than us and possibly was concerned about leaving a trace of its existence, or of *ii)* a very advanced, old and stabilized civilization who built macro-engineered structures, either for communication purposes or completely different needs.

Acknowledgements: The author thanks the two referees, Jason W. Barnes and another anonymous referee, for their very constructive comments on the manuscript.